\documentclass{aastex}          %% The manuscript based on AASTeX v5.x
\usepackage{spr-astr-addons}    %% mimicing ASTR journal style
\usepackage{url}

\RequirePackage{color}
\newcommand{\be}{\begin{equation}}
\newcommand{\ee}{\end{equation}}
\newcommand{\bea}{\begin{eqnarray}}
\newcommand{\eea}{\end{eqnarray}}

\begin{document}
%% Article title
%
\title{Acceleration of the Universe in f(R) Gravity Models}

%% Running heads
\shorttitle{<Short article title>}
\shortauthors{<Ankan Mukherjee and Narayan Banerjee>}

%% Author and Affilations
\author{Ankan Mukherjee\altaffilmark{}} 
\affil{Department of Physical Sciences,~~\\Indian Institute of Science Education and Research Kolkata,~~\\Mohanpur Campus, Mohanpur, West Bengal -741252,India. \\ ankan$\_$ju@iiserkol.ac.in}
\and 
\author{Narayan Banerjee\altaffilmark{}}
\affil{Department of Physical Sciences,~~\\Indian Institute of Science Education and Research Kolkata,~~\\Mohanpur Campus, Mohanpur, West Bengal -741252,India. \\ narayan@iiserkol.ac.in}

%% Abstract
\begin{abstract}
A general formalism for the investigation of the late time dynamics of the universe for any analytic f(R) gravity model, along with a cold dark matter, has been discussed in the present work. The formalism is then elucidated with two examples. The values of the parameters of the models are chosen in such a way that they are consistent with the basic observational requirement.
\end{abstract}



\section{Introduction}   %\label{Introduction}
Cosmology has indeed undergone a dramatic change over the past couple of decades. The availability of high precision data regarding our universe and its indication towards an expanding universe with an accelerated rate invoked all sorts of modifications of Einstein's equations. That the acceleration must have set in a not too distant past is a theoretical requirement as well as has been observationally supported. There are excellent reviews regarding the accelerated expansion\citep{sahni2000,paddy2003,copeland2006}.

\par The theoretical investigation towards finding a viable option which can drive this acceleration is done in two distinct ways. One way is to modify the matter sector by adding an exotic field giving rise to an effective negative pressure. The most talked about agent capable of doing this is certainly the cosmological constant $\Lambda$. A scalar field with some potential, known as the quintessence matter, is also amongst the most favourite candidates as a "dark energy", the agent driving this alleged expansion. We refer to the reviews \citep{sahni2000,paddy2003,copeland2006}  and the references therein. 

\par The second option is to look for a theory of gravity where the Einstein-Hilbert action is modified. One way to do that is to consider a scalar field nonminimally coupled to the geometry sector \citep{nbdp2001a,nbdp2001b,ss2001,bertolami2000,elizalde2004,onemliwood2002,onemliwood2004,brunier2005,dasbaneree2006} or to the matter sector \citep{khoury2004a,khoury2004b,mota2004a,mota2004b,dass2006,nb2010}  or both \citep{dasbaneree2008}. The other popular way is to use an analytic function $f=f(R)$ in place of $R$ in the action, where $R$ is the Ricci scalar \citep{capo2003,brofra2004,nojiri2003,dolgov2003,carr2004,dsnbnd2006}. It had already been noted that $f(R) \propto R^2$ kind of theories could successfully generate inflationary universe scenario for the early universe \citep{starobaa1980,kerner1982,duruisseau1986}. As the curvature $R$ decreases with time, inverse power of $R$ in an $f(R)$ theory might be expected to generate a late time acceleration. For a detailed description of $f(R)$ theories and their application  in cosmological models, we refer to some recent reviews \citep{sotiriou2010,felice2010,nojiriod2011}. The $f(R)$ gravity models, available in the literature mostly deal with only the present acceleration and hardly talk about the smooth transition from a decelerated to an accelerated regime. There are anyway a few investigations regarding this signature flip in the deceleration parameter $q$. For instance, we refer to \cite{nojiodin2006} and  \cite{nojiodinstef2006}. Nojiri and Odintsov \citep{nojiri2007a} also reconstructed an $f(R)$ gravity model from a $\Lambda$CDM one.

\par Das, Banerjee and Dadhich \citep{dsnbnd2006} indeed discussed models that show such a smooth transition analytically, but the models do not contain matter. However, it has been shown that, along with a matter field, any such model could behave in quite a different manner\citep{amendola1,amendola2}. Some models where the Ricci scalar is non-minimally coupled to the matter sector are also there in literature \citep{thakur2011}. Some of the modifications of the geometry sector also involves a function of $G$, the Gauss-Bonnet scalar. Nojiri and Odintsov discussed $f(G)$ and $f(R,G)$ models in connection with the recent accelerated expansion of the universe \citep{nojiri2007d}. A reconstruction of $f(R)$ gravity model can be found in reference\citep{nojiri2007e}. As the $\Lambda$CDM model does well regarding the fits with the observational data, there are attempts to distinguish between models that mimic the $\Lambda$CDM model and those which do not. For example, the models given by Hu and Sawicki \citep{hu2007} and Starobinsky \citep{starob2007} are distinct from the $\Lambda$CDM, whereas those given by He and Wang \citep{he2013} and Dunsby et al \citep{dunsby2010} are consistent with that. Constraining the model parameters of an $f(R)$ gravity model has been discussed recently by Nojiri and Odintsov \citep{nojiri2007b, nojiri2007c},  Giron$\acute{e}s$ {\it et al} \citep{giro2010} and Basilakos {\it et al} \citep{basila2013}.

\par One general problem with $f(R)$ theories is that they either give an early inflation or a late time acceleration. There have been recent attempts to find some form of $f(R)$ which would yield accelerated expansion in two phases, one in an early epoch and the other in the late stage of evolution. Cognola {\it et al} \citep{cognola2009} and Elizalde {\it et at} \citep{elizalde2011} made such attempts with an $f(R)$ which is an exponential function of $R$. Nojiri and Odintsov \citep{odin2003} made an attempt to unify the two phases of accelerated expansion in the realm of a single $f(R)$ gravity model by combining positive and negative powers of $R$. Possible impacts of the existence of nonlinear terms involving $R$ in the action on the structure formation has been discussed by Thakur and Sen\citep{anjan2013}.

\par In the present work, a straightforward way to facilitate the investigation of the dynamics of the universe is discussed. The net conservation equation results from the contracted Bianchi identity. We assume that the matter content obeys its own conservation, which, in tandem with the net conservation equation yields an equation for the contribution from the geometry sector to the evolution of the universe. This equation is a second order differential equation in the Hubble parameter $H$.  The equation is highly nonlinear and it is difficult to get an analytic solution. However, with proper boundary conditions, one can plot the relevant cosmological parameters like the deceleration parameter $q$, the effective equation of state parameter $w_{eff}$ such that the qualitative behaviour of the model is understood. We deal with two simple examples to elucidate the method, a two-parameter model ($f(R) \propto (\lambda +R)^{n}$) and a one-parameter model ($f(R) \propto exp(\alpha R)$).

\section{$f(R)$ gravity and the conservation equation}

The generalized Einstein-Hilbert action for $f(R)$ gravity is
\be
{\cal A}=\int \Bigg[\frac{1}{16\pi G}f(R)+{\cal L}_m\Bigg]\sqrt{-g}d^4x,
\label{action}
\ee
where $R$ is replaced by $f(R)$ in the Einstein-Hilbert action, $f(R)$ being an analytic function of $R$. Here ${\cal L}_m$ is the usual matter field Lagrangian. A variation of this action, with respect to the metric, yields the field equations as
\be
f'(R)R_{\mu\nu}-\nabla_\mu\nabla_\nu f'(R)+\big[\Box f'(R)-\frac{1}{2}f(R)\big]g_{\mu\nu}=T^{(m)}_{\mu\nu},
\label{fRfieldeq}
\ee
where a prime indicates differentiation with respect to the Ricci scalar $R$ and $T^{(m)}_{\mu\nu}$ represents the contribution to the energy momentum tensor from matter fields with a choice of unit as $8\pi G=1$. This variation is popularly dubbed as the metric $f(R)$ gravity as opposed to the Palatini formulation where the variation is carried out with respect to both the metric and the affine connections. 

\par The present endeavour is to study the dynamics of the universe in the background of the spatially flat FRW metric, which  is written as
\be
ds^2=dt^2-a^2(t)[dr^2+r^2d\theta^2+r^2sin^2\theta d\phi^2],
\label{metric}
\ee
where $a(t)$ is the scale factor. The field equations take the form
\be
3\frac{\dot{a}^2}{a^2}=\frac{\rho_m}{f'}+\frac{1}{f'}\Big[\frac{1}{2}(f-Rf')-3\dot{R}f''\frac{\dot{a}}{a}\Big],
\label{frd1}
\ee  
\be
2\frac{\ddot{a}}{a}+\frac{\dot{a}^2}{a^2}=-\frac{1}{f'}\Big[\ddot{R}f''+\dot{R}^2f'''+2\dot{R}f''\frac{\dot{a}}{a}-\frac{1}{2}(f-Rf')\Big],
\label{frd2}
\ee
where dots are the derivatives with respect to cosmic time and a prime denotes derivative with respect to $R$. Also, $\rho_m$ is the dark matter density and, consistent with the cold dark matter, the corresponding pressure $p_m$ is taken to be zero.

\par We write 
\be
\Big[\frac{1}{2}(f-Rf')-3\dot{R}f''\frac{\dot{a}}{a}\Big]=\rho_c,
\label{rhoc}
\ee
and 
\be
\Big[\ddot{R}f''+\dot{R}^2f'''+2\dot{R}f''\frac{\dot{a}}{a}-\frac{1}{2}(f-Rf')\Big]=p_c,
\label{pc}
\ee
as $\rho_c$ and $p_c$ determine the contribution by the curvature to the density and pressures sectors respectively. For $f(R) = R$, both $\rho_c$ and $p_c$ would vanish as expected. In terms of the Hubble parameter ($H=\frac{\dot{a}}{a}$), the modified field equations (\ref{frd1}) and (\ref{frd2}) will read as 
\be
3H^2=\frac{\rho_m+\rho_c}{f'},
\label{frdhubble1}
\ee
and
\be
2\dot{H}+3H^2=-\frac{p_c}{f'},
\label{frdhubble2}
\ee
respectively.

Finally, the contracted Biacchi identity will yield
\be
\frac{d}{dt}\Big(\frac{\rho_m+\rho_c}{f'}\Big)+3H\Big(\frac{\rho_m+\rho_c+p_c}{f'}\Big)=0.
\label{BianchiID}
\ee
Considering the matter conservation, i.e., $\dot{\rho_m}+3H\rho_m=0$, is satisfied independently, equation (\ref{BianchiID}) will be simplified to the form
\be
\frac{d}{dt}\Big(\frac{\rho_c}{f'}\Big)+3H\Big(\frac{\rho_c+p_c}{f'}\Big)=\rho_m\Big(\frac{\dot{R}f''}{f'}\Big).
\label{BianchiID2}
\ee
This equation can be written in the form
\be
18\frac{f''}{f'}H(\ddot{H}+4H\dot{H})+3(\dot{H}+H^2)+\frac{f}{2f'}+\frac{\rho_m}{f'}=0,
\label{conservationequ}
\ee
where use has been made of the expressions $R=-6(\dot{H}+2H^2)$ and $\dot{R}=-6(\ddot{H}+4H\dot{H})$. It is important to note that equation (\ref{BianchiID}) yields equation (\ref{conservationequ}) under the condition $f'' \neq 0$. So one cannot arrive at the corresponding equation for $f(R)=R$, the usual Einstein-Hilbert action. Now we write the equation (\ref{conservationequ}) with the redshift $z$ (given by $1+z=\frac{a_0}{a}$) as the argument. The equation now looks like
\be
\begin{split}
\frac{d^2H}{dz^2}=\frac{3}{(1+z)}\frac{dH}{dz}-\frac{1}{H}\Bigg(\frac{dH}{dz}\Bigg)^2  \\
-\frac{3f'\Big(H^2-(1+z)H\frac{dH}{dz}\Big)+\frac{f}{2}+\rho_{m0}(1+z)^3}{18(1+z)^2H^3f''}.
\label{conservationequ2}
\end{split}
\ee 
This is the key equation in our attempt to study the dynamics of the universe in $f(R)$ gravity models. Though the equation is highly non-linear, we can at least investigate the the redshift dependence of the Hubble parameter $H(z)$ and other important parameters numerically when the form of $f(R)$ is given. 
\par Now from equation (\ref{frdhubble1}), the present matter density $\rho_{m0}$ can be expressed as 
\be
\rho_{m0}=3H_0^2f'_o-\rho_{c0}.
\ee
where a subscript 0 indicates the values of the functions at the present epoch, namely at $z=0$. Now $R=-6H^2(1-q)$ and $\dot{R}=-6H^3(j-q-2)$ where $q=-\frac{\ddot{a}}{aH^{2}}$ is the deceleration  and $j=\frac{\ddot{a}}{a^3 H^3}$ is the jerk parameter. Hence present value of Ricci scalar and its derivative can be estimated from the knowledge of present deceleration parameter $q_0$ and jerk $j_0$. We scale $H$ as $\frac{H}{H_0}$ so that the present value of Hubble parameter $H_0$ is unity. A simple dimensional consideration shows that this can be done without any loss of generality in the equation (\ref{conservationequ2}) by dividing both sides by $H_0$. There are observational estimates for the parameters $q_0$ and $j_0$. In the present work, we pick up the relevant values from the work of Rapetti {\it et al}\citep{rapetti2007}. The relevant values are $q_0=-0.81\pm0.14$ and $j_0=2.16_{-0.75}^{+0.81}$.

\section{$f(R)$ gravity in a spatially flat FRW universe}

With a functional form of $f(R)$, equation (\ref{conservationequ2}), a second order differential equation in $H(z)$, can be numerically integrated. In this work, two  $f(R)$ gravity models have been discussed. The aim is to find the parameters of the models that would be in agreement with the observed values of the relevant cosmolgical parameters, namely the deceleration parameter $q$ and the effective equation of state parameter $w_{eff}$ given by $w_{eff}=\frac{p_c}{\rho_c + \rho_m}$.

\subsection{CaseI: $f(R)= {\lambda_0}(\lambda+R)^n$}
We choose $f(R)={\lambda_0}(\lambda+R)^n$, where $\lambda_0$, $\lambda$ and $n$ are constants and they actually are the model parameters.As $f(R)$ should have the dimension of $R$, the constant $\lambda_0$ is there to take care of the dimension. In all subsequent discussion, the value of the constant $\lambda_0$ is taken to be unity. From equation (13), the numerical plots of deceleration parameter $q(z)$ and the effective equation of state parameter $w_{eff}(z)$ are obtained using the present values $q_0=-0.81\pm0.14$ and $j_0=2.16_{-0.75}^{+0.81}$ as mentioned in the previous section. There are two parameters in the model, namely $\lambda$ and $n$. The plots have been generated taking four sets of values of these two parameters. Each set has been adjusted in such a way that the recent acceleration starts around $z=0.5$. Figures 1-4 show these plots. The present value of the parameter $w_{eff}$ in all cases is between -0.8 to -1.0. This also in conformity with the observational estimate. It deserves mention that most of the examples of the power law type $f(R)$ gravity models leading to present acceleration involves some negative power for the Ricci scalar $R$ in the action. But in all the examples in this work, $n$ is positive and so there is no singularity in $f(R)$ for $R$ going to zero.

%%%%%%%%%%%%%%%%%%%%%%%%%%%%%%%%%%%%%%%%%%%%%%%%%%%%%%%%%%%%%%%%%%%%
\begin{figure}[hbtp]
\centering
\includegraphics[width=\columnwidth]{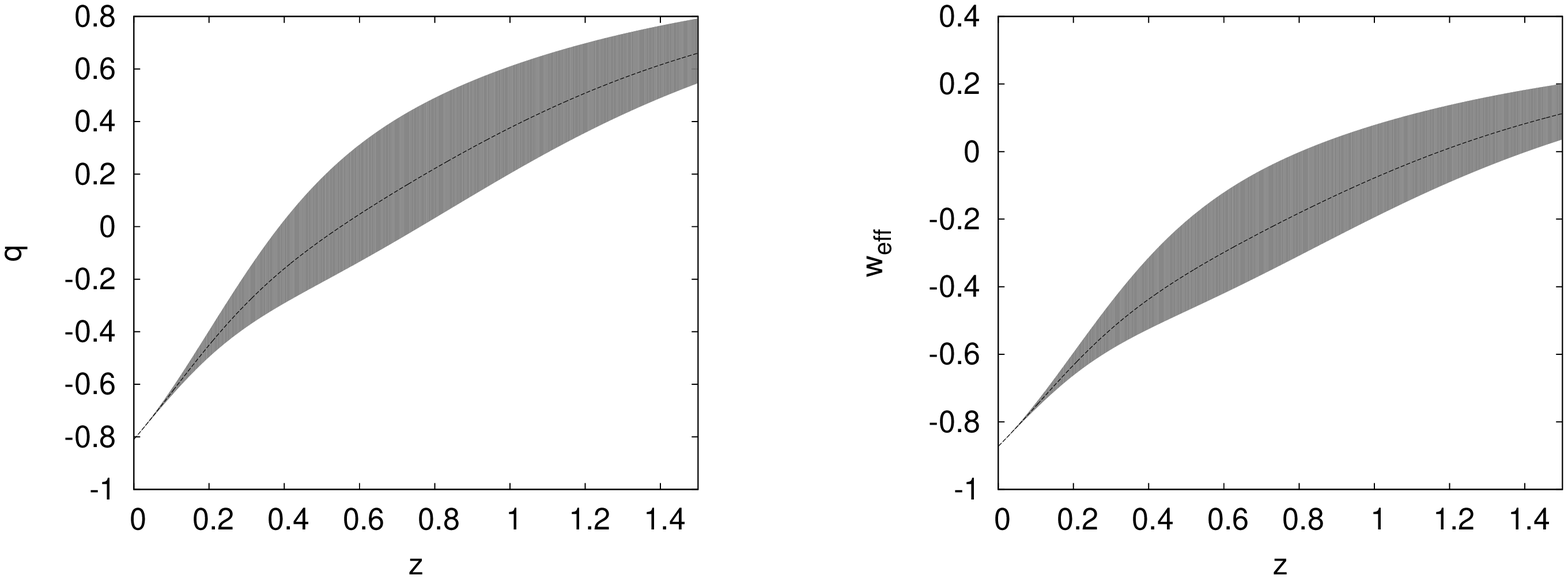}
\caption{\small Plots of deceleration parameter $q$ (left panel) and effective equation of state $w_{eff}$ (right panel) against redshift $z$ for $f(R)=\lambda_0(\lambda+R)^n$ with $\lambda_0=1$, $n=0.5$ and $\lambda=13.5\pm0.5$. The central dark line is for $\lambda=13.5$ and  $\lambda=13.0$ is the upper and $\lambda=14.0$ is the lower bounds of the plots. }
\end{figure}
%%%%%%%%%%%%%%%%%%%%%%%%%%%%%%%%%%%%%%%%%%%%%%%%%%%%%%%%%%%%%%%%%%%%%%%%%%

%%%%%%%%%%%%%%%%%%%%%%%%%%%%%%%%%%%%%%%%%%%%%%%%%%%%%%%%%%%%%%%%%%%%%%%%%%
\begin{figure}[hbtp]
%\begin{center}
\centering
\includegraphics[width=\columnwidth]{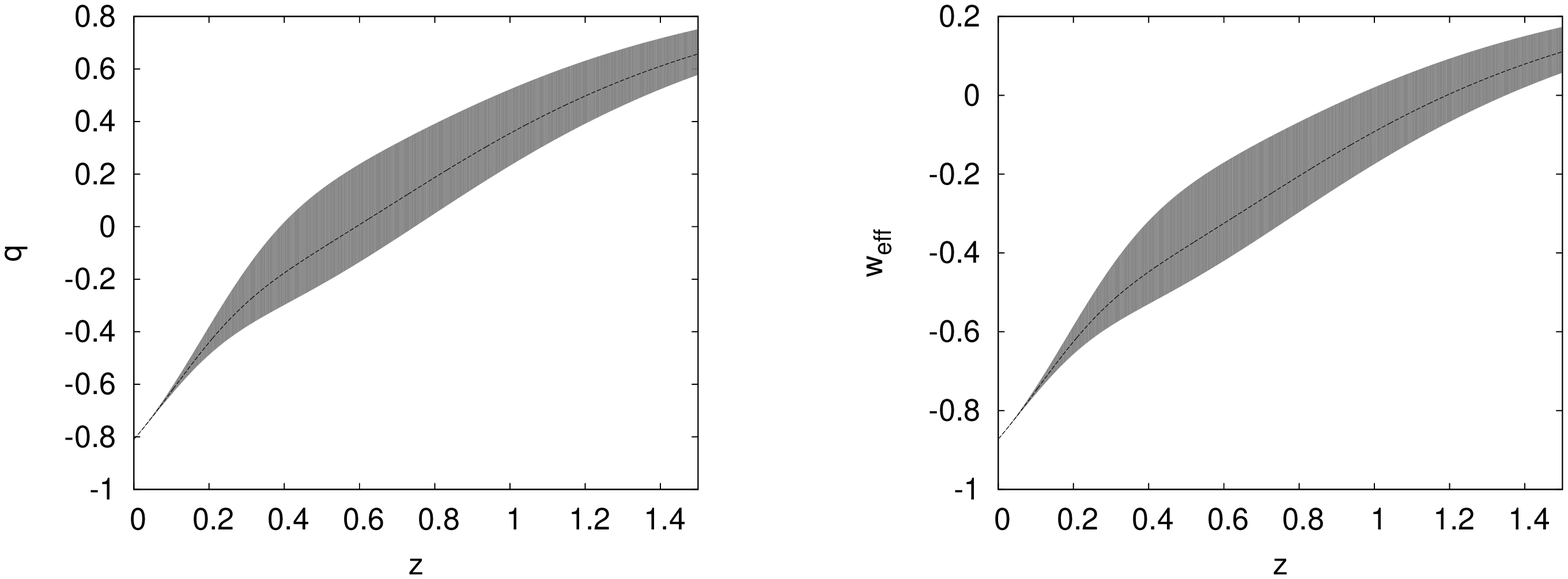}
%\end{center}
\caption{\small Plots of deceleration parameter $q$ (left panel) and effective equation of state $w_{eff}$ (right panel) against redshift $z$ for $f(R)=\lambda_0(\lambda+R)^n$ with $\lambda_0=1$, $n=0.1$ and $\lambda=13.0\pm0.25$. The central dark line is for $\lambda=13.0$ and  $\lambda=12.75$ is the upper and $\lambda=13.25$ is the lower bounds of the plots.}
\end{figure}
%%%%%%%%%%%%%%%%%%%%%%%%%%%%%%%%%%%%%%%%%%%%%%%%%%%%%%%%%%%%%%%%%%%%%%%%%%%%

%%%%%%%%%%%%%%%%%%%%%%%%%%%%%%%%%%%%%%%%%%%%%%%%%%%%%%%%%%%%%%%%%%%%%%%%%%%%
\begin{figure}[hbtp]
%\begin{center}
\centering
\includegraphics[width=\columnwidth]{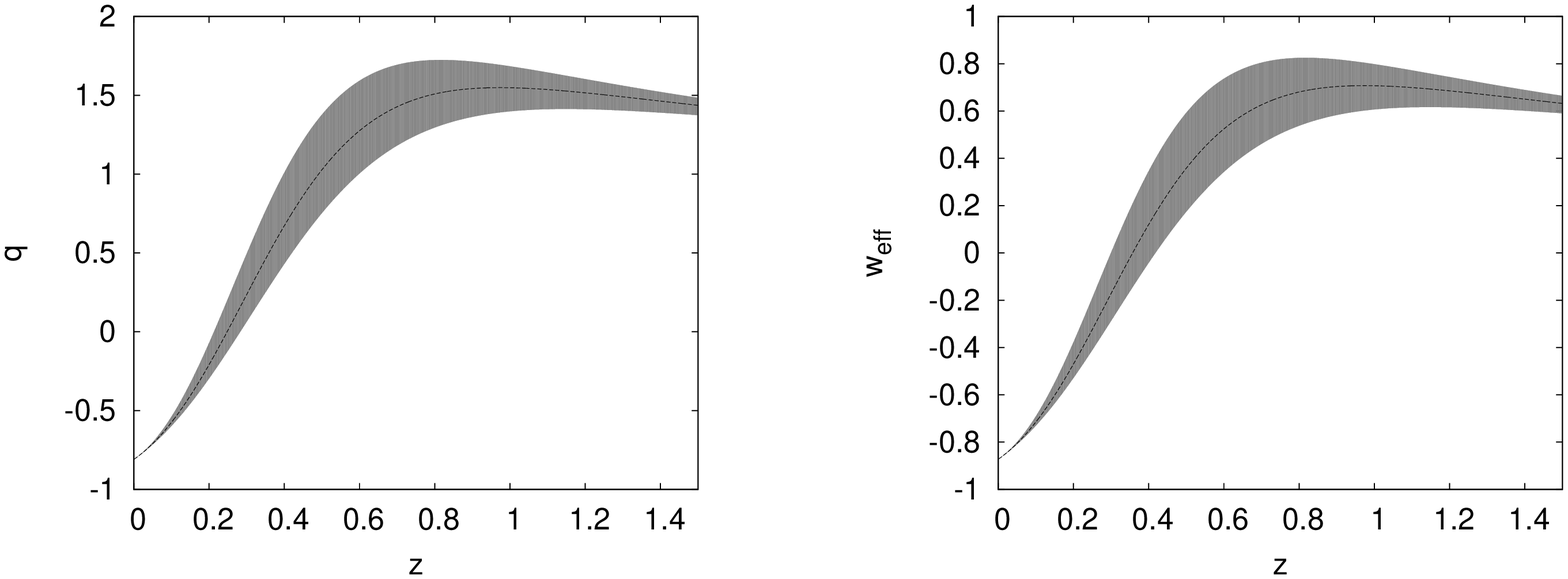}
%\end{center}
\caption{\small Plots of deceleration parameter $q$ (left panel) and effective equation of state $w_{eff}$ (right panel) against redshift $z$ for $f(R)=\lambda_0(\lambda+R)^n$ with $\lambda_0=1$,  $n=1.5$ and $\lambda=16.0\pm3.0$. The central dark line is for $\lambda=16.0$ and  $\lambda=19.0$ is the upper and $\lambda=13.0$ is the lower bounds of the plots.}
\end{figure}
%%%%%%%%%%%%%%%%%%%%%%%%%%%%%%%%%%%%%%%%%%%%%%%%%%%%%%%%%%%%%%%%%%%%%%%%%%%%%%%%

%%%%%%%%%%%%%%%%%%%%%%%%%%%%%%%%%%%%%%%%%%%%%%%%%%%%%%%%%%%%%%%%%%%%%%%%%%%%%%%%
\begin{figure}[hbtp]
%\begin{center}
\centering
\includegraphics[width=\columnwidth]{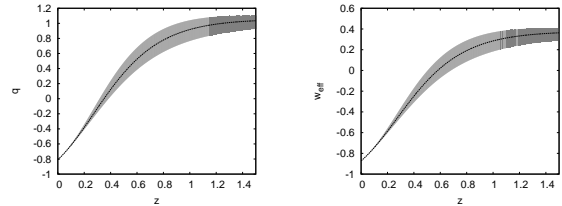}
%\end{center}
\caption{\small Plots of deceleration parameter $q$ (left panel) and effective equation of state $w_{eff}$ (right panel) against redshift $z$ for $f(R)=\lambda_0(\lambda+R)^n$ with $\lambda_0=1$, $n=2.0$.  The central dark line is for $\lambda=12.0$ and  $\lambda=15.0$ is the upper and $\lambda=1.0$ is the lower bounds of the plots.}
\end{figure}
%%%%%%%%%%%%%%%%%%%%%%%%%%%%%%%%%%%%%%%%%%%%%%%%%%%%%%%%%%%%%%%%%%%%%%%%%%%%%

\subsection{Case II: $f(R)=R_0exp(\alpha R)$}
This exponential form of $f(R)$ had already been discussed in \citep{dsnbnd2006}. However, that was done with no matter content of the universe. In the present work, the same single parameter exponential form of $f(R)$ has been introduced along with the matter content. Like the previous example, the constant $R_0$ takes care of the dimensional requirement and is chosen to be unity in the subsequent discussion. The numerical plots for the deceleration parameter $q$ and the effective equation of state parameter $w_{eff}$ are obtained for a range of values of $\alpha$ (between 0.5 and 15.0) with the similar boundary conditions used for the previous model. The range of values of $\alpha$ are chosen so as to get the signature flip in $q$ close to $z=0.5$. Figure 5 clearly shows that this model also successfully generates late time acceleration accompanied by the decelerated expansion era that prevailed earlier. The central curve is for $\alpha = 1.5$ for both of $q$ and $w_{eff}$. If the valu
 e of $\alpha$ is raised to 15, the lower curve is obtained. But almost similar amount of deviation is seen for the higher curve for a much smaller variation of the parameter. The upper curve is obtained when $\alpha$ is changed to 0.5. So the amount of acceleration is much more sensitive to a decrease of the parameter of the model.

%%%%%%%%%%%%%%%%%%%%%%%%%%%%%%%%%%%%%%%%%%%%%%%%%%%%%%%%%%%%%%%%%%%%%%  
\begin{figure}[hbtp]
\begin{center}
\includegraphics[width=\columnwidth]{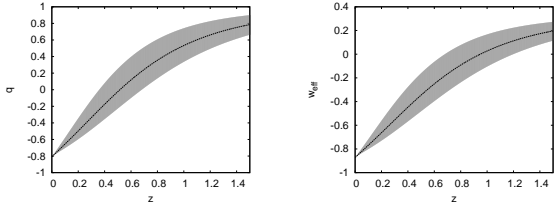}
\end{center}
\caption{\small Plots of deceleration parameter $q$ (left panel) and effective equation of state $w_{eff}$ (right panel) against redshift $z$ for $f(R)=R_0exp(\alpha R)$.  The central dark line is for $R_0=1$, $\alpha=1.5$ and  $\alpha=0.5$ is the upper and $\alpha=15.0$ is the lower bounds of the plots.}
\end{figure}
%%%%%%%%%%%%%%%%%%%%%%%%%%%%%%%%%%%%%%%%%%%%%%%

\vskip 2.5 cm

\section{Discussion}
A straightforward way for the discussion of the dynamics of the much talked about $f(R)$ gravity models along with a cold dark matter content has been presented in this work.As it has been shown that an $f(R)$ gravity model could behave in a dramatically different manner in the presence of matter\citep{amendola1,amendola2}, it is imperative that the models are discussed in the presence of matter. Equation (\ref {conservationequ2}) sets up a basic framework for that. Both the models presented here work well  in the presence of matter.
\par Two examples have been worked out, one of them, namely the case I is apparently new, and the second case has already been discussed, although without the requisite matter content. The parameters of the model are reconstructed from the observational values of some cosmological parameters. However, no rigorous statistical analysis has been employed for the estimation of the model parameters. 

\par It deserves mention at  this stage that the two models presented here do not have the same degree of stability. If a quantity, $m^2=\frac{1}{3}\Big[\frac{f'(R)}{f''(R)}-R\Big]$, is defined at $R=R_0$, the present value of the Ricci curvature, one can show that $m^2<0$ leads to a tachyonic instability \citep{nojiri2007b}. The second model of the present work (section 3.2) has this instability. Our first model  $f(R)=\lambda_0(\lambda+R)^n$, on the other hand, passes this fitness test. 
\par The primary motivation is to set up a general framework, but both the examples discussed can produce a signature flip at the right epoch and can reproduce the total effective equation of state parameter $w_{eff}$ close to its expected present value. This basic observational requirement is met for actually quite a wide range of the model parameters. 
\par It also deserves mention that according to the criterion discussed by Basilakos\citep{basila2013}, none of the two models presented here would actually converge to the $\Lambda CDM$ model. The first example would do that only for the trivial case of $n=1$.

%\smallskip 
\noindent
%{\setlength{\parindent}{0cm}
\nocite{*}
 \bibliographystyle{spr-mp-nameyear-cnd}  %% BibTeX style
 \bibliography{biblio}                %% BibTeX data

\end{document}